\documentclass[%
reprint, 
superscriptaddress,
 amsmath,amssymb,
 aps, 
floatfix,
]{revtex4-2}

\usepackage[english]{babel}
\DeclareRobustCommand{\selectlanguage}[1]{}
\usepackage[T1]{fontenc}
\usepackage[utf8]{inputenc}
\usepackage{graphicx}
\usepackage{dcolumn}
\usepackage{bm}
\usepackage[thinc]{esdiff}
\usepackage{xcolor}

\begin{document}

\preprint{APS/123-QED}

\title{\textbf{High fidelity qubit control in a natural Si-MOS quantum dot using a 300 mm silicon on insulator wafer}
}%

\author{Xander Peetroons}
\email{xprp2@cam.ac.uk}
\affiliation{Hitachi Cambridge Laboratory, J. J. Thomson Avenue, Cambridge CB3 0US, United Kingdom}
\affiliation{Cavendish Laboratory, Department of Physics, University of Cambridge,  Cambridge CB3 0US, United Kingdom}
 
\author{Xunyao Luo}
\affiliation{Hitachi Cambridge Laboratory, J. J. Thomson Avenue, Cambridge CB3 0US, United Kingdom}
\affiliation{Cavendish Laboratory, Department of Physics, University of Cambridge,  Cambridge CB3 0US, United Kingdom}
\author{Tsung-Yeh Yang}
\affiliation{Hitachi Cambridge Laboratory, J. J. Thomson Avenue, Cambridge CB3 0US, United Kingdom}
\author{Normann Mertig}
\affiliation{Hitachi Cambridge Laboratory, J. J. Thomson Avenue, Cambridge CB3 0US, United Kingdom}
\author{Sofie Beyne}
\affiliation{IMEC, Leuven, Belgium}
\author{Julien Jussot}
\affiliation{IMEC, Leuven, Belgium}
\author{Yosuke Shimura}
\affiliation{IMEC, Leuven, Belgium}

\author{Clement Godfrin}
\affiliation{IMEC, Leuven, Belgium}
\author{Bart Raes}
\affiliation{IMEC, Leuven, Belgium}
\author{Ruoyu Li}
\affiliation{IMEC, Leuven, Belgium}
\author{Roger Loo}
\affiliation{IMEC, Leuven, Belgium}
\affiliation{Dept. of Solid-State Sciences, Ghent University, Krijgslaan 281, building S1, 9000 Ghent, Belgium}
\author{Sylvain Baudot}
\affiliation{IMEC, Leuven, Belgium}
\author{Stefan Kubicek}
\affiliation{IMEC, Leuven, Belgium}
\author{Shuchi Kaushik}
\affiliation{IMEC, Leuven, Belgium}
\author{Danny Wan}
\affiliation{IMEC, Leuven, Belgium}
\author{Takeru Utsugi }
\affiliation{Research and Development Group, Hitachi, Ltd., Kokubunji, Tokyo 185-8601, Japan}
\author{Takuma Kuno }
\affiliation{Research and Development Group, Hitachi, Ltd., Kokubunji, Tokyo 185-8601, Japan}
\author{Noriyuki Lee}
\affiliation{Research and Development Group, Hitachi, Ltd., Kokubunji, Tokyo 185-8601, Japan}
\author{Itaru Yanagi}
\affiliation{Research and Development Group, Hitachi, Ltd., Kokubunji, Tokyo 185-8601, Japan}
\author{Toshiyuki Mine}
\affiliation{Research and Development Group, Hitachi, Ltd., Kokubunji, Tokyo 185-8601, Japan}
\author{Satoshi Muraoka}
\affiliation{Research and Development Group, Hitachi, Ltd., Kokubunji, Tokyo 185-8601, Japan}
\author{Shinichi Saito}
\affiliation{Research and Development Group, Hitachi, Ltd., Kokubunji, Tokyo 185-8601, Japan}
\author{Digh Hisamoto}
\affiliation{Research and Development Group, Hitachi, Ltd., Kokubunji, Tokyo 185-8601, Japan}
\author{Ryuta Tsuchiya}
\affiliation{Research and Development Group, Hitachi, Ltd., Kokubunji, Tokyo 185-8601, Japan}
\author{Hiroyuki Mizuno}
\affiliation{Research and Development Group, Hitachi, Ltd., Kokubunji, Tokyo 185-8601, Japan}
\author{Kristiaan De Greve}
\affiliation{IMEC, Leuven, Belgium}
\affiliation{Proximus chair in quantum science and technology, KU Leuven, Leuven, Belgium}
\affiliation{Dept. of electrical engineering (ESAT-MNS), KU Leuven, Leuven, Belgium}
\author{Charles Smith}
\affiliation{Hitachi Cambridge Laboratory, J. J. Thomson Avenue, Cambridge CB3 0US, United Kingdom}
\affiliation{Cavendish Laboratory, Department of Physics, University of Cambridge,  Cambridge CB3 0US, United Kingdom}
\author{Helena Knowles}
\affiliation{Cavendish Laboratory, Department of Physics, University of Cambridge,  Cambridge CB3 0US, United Kingdom}
\author{Andrew Ramsay}
\affiliation{Hitachi Cambridge Laboratory, J. J. Thomson Avenue, Cambridge CB3 0US, United Kingdom}


\date{\today}

\begin{abstract}

We demonstrate high-fidelity single qubit control in a natural Si-MOS quantum dot fabricated in an industrial 300 mm wafer process on a silicon on insulator (SOI) wafer using electron spin resonance. A relatively high optimal Rabi frequency of 5 MHz is achieved, dynamically decoupling the electron spin from its \(^{29}\mathrm{Si}\) environment. Tracking the qubit frequency reduces the impact of low frequency noise in the qubit frequency and improves the \(T^{Rabi}\) from 7 to 11 \(\mu s\) at a Rabi frequency of 5 MHz, resulting in Q-factors exceeding 50. Randomized benchmarking returns an average single gate control fidelity of $99.5\pm0.3\%$. As a result of pulse-area calibration, this fidelity is limited by the Rabi Q-factor. These results show that a fast Rabi frequency, low charge noise, and a feedback protocol enable high fidelity in these Si-MOS devices, despite the low-frequency magnetic noise.
\end{abstract}

\maketitle

\section{Introduction}

Spin qubits in semiconductor quantum dots \cite{PhysRevA.57.120}, are a widely studied pathway towards realizing large-scale quantum computers. \cite{Kunne2024-wk} In terms of industrial scale fabrication, and integration with control electronics silicon spin qubits have a clear advantage, \cite{Gonzalez-Zalba2021-kg, Zwerver2022-mi, Veldhorst2017-ht} and exhibit long coherence times compared to other gate-defined solid-state qubits. \cite{Veldhorst2014-la} To enable large-scale quantum processors it is important to increase the control fidelities of single- and two-qubit gates \cite{Steinacker2024-xp}. Additionally, surface code error correction requires a maximum error rate of 1\% to be viable. \cite{PhysRevA.86.032324, Wang2011-da}

Recent demonstrations have achieved gate fidelities exceeding 99.9\% in isotopically purified $^{28}$Si/SiGe platforms \cite{Mills2022-yi, Philips2022-oo, Noiri2022-ou} and in isotopically purified $^{28}$Si-MOS structures \cite{Yang2019-rz, Gilbert2023-hh}. However, the control fidelities in isotopically natural silicon remain low, due to the noisy nuclear spin environment of the qubit. Through hyperfine interactions with the spin-half \(^{29}\mathrm{Si}\) nuclei, a single electron spin qubit experiences fluctuations in an effective magnetic field. These hyperfine interactions are especially detrimental in n-type natural silicon devices. \cite{Cvitkovich2024-nq, Kawakami2016-gk} Previously, gate fidelities above the fault-tolerant threshold in natural Si-MOS have only been shown in a p-type Si-MOS device using electric dipole spin resonance (EDSR) reaching 99.8\%, where the hyperfine interactions are suppressed by the p-wave nature of holes. \cite{Vorreiter2025-lv} A single qubit gate fidelity of 99.1\% in a natural Si-MOS device can also be achieved using a phase-modulated drive. \cite{Kuno2025-cn} Charge noise also causes fluctuations in the qubit frequency via spin-orbit interaction, making it difficult to fully identify the source of noise. \cite{Rojas-Arias2024-zt} Si/SiGe heterostructures achieve reduced levels of charge noise by positioning the qubit further from the defects at the oxide interface. This is at the cost of a reduced lever arm, a fabrication process that is less compatible with standard 300 mm wafer practices, and smaller valley splittings. \cite{Neyens2024-vn} Single gate fidelities in natural silicon in Si/SiGe platforms with EDSR have reached 99.2\% \cite{Wang2024-gd} and 99.6\%. \cite{Takeda2016-vb} Despite the noise-susceptible interface of Si-MOS, compared to Si/SiGe devices, low levels of charge noise have been achieved in a 300 mm wafer line. \cite{Elsayed2024-io} 

Fluctuations in the qubit frequency cause decoherence, and real-time feedback to track the electron spin resonance (ESR) helps to attain high control fidelity. \cite{PhysRevA.49.2133, Park2025-dl, Vepsalainen2022-yu} Since the feedback compensates noise fluctuations with a correlation time slower than the time to perform a calibration, the speed of the calibration is key. Similarly, real-time two-axis control feedback was recently demonstrated on a singlet-triplet qubit in GaAs. \cite{Berritta2024-zw} Several feedback protocols have also been developed to track different qubit parameters in real-time. \cite{Dumoulin_Stuyck2024-ql} Additionally, fine calibration of the control pulse lengths and amplitudes are necessary to reach high control fidelities. \cite{Smith2024-dv, Dumoulin_Stuyck2024-ql}

\begin{figure*}[ht]
\includegraphics[width=0.8\linewidth]{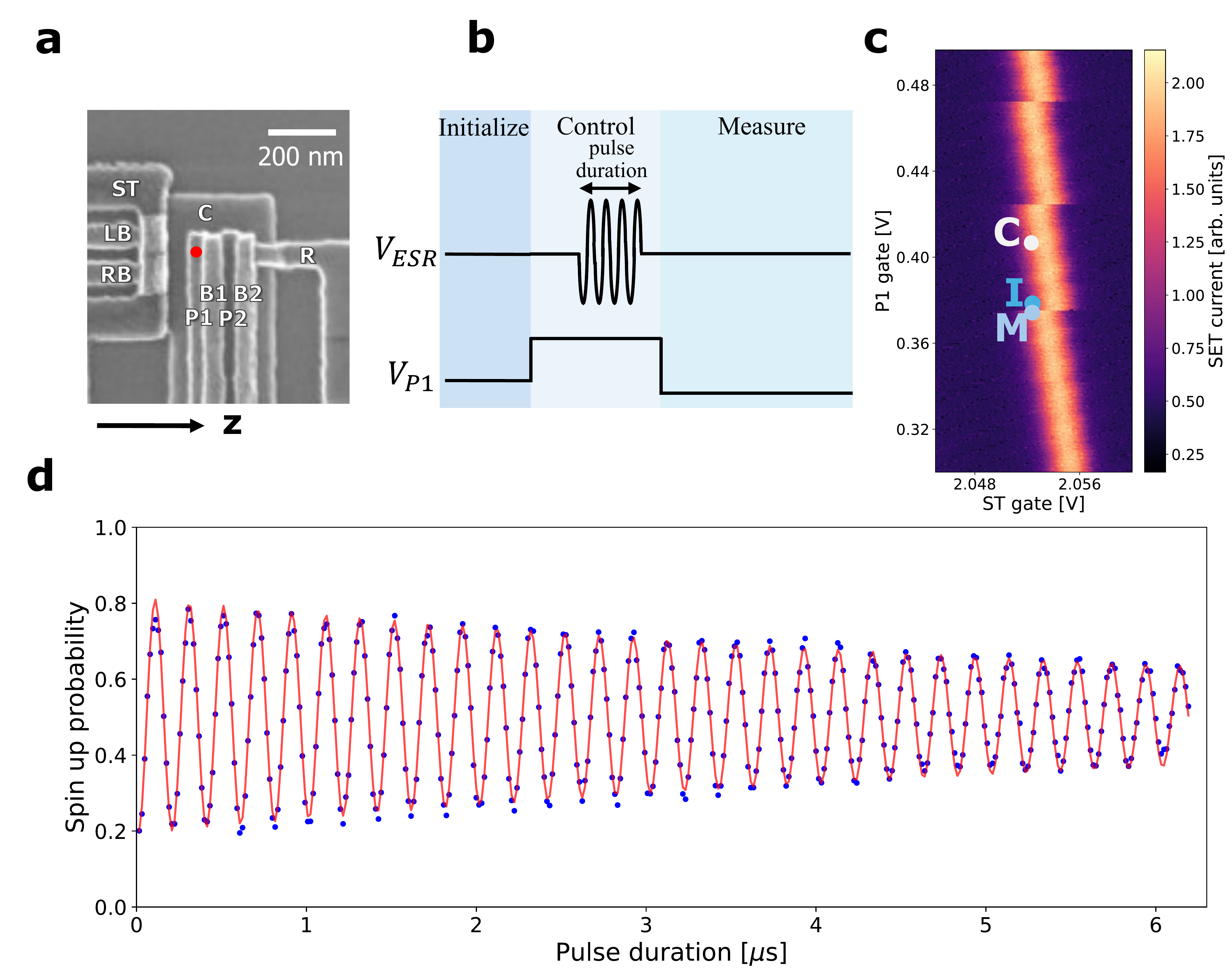}
\caption{\label{fig:fig1} \textbf{Single qubit control} a) Scanning electron micrograph of the device with the quantum dot indicated by the red dot, next to the single electron transistor, b) Pulse sequence for single qubit control, c) Charge stability diagram of the P1 quantum dot indicating the initialize (I), control (C), and measure (M) positions for the ST and P1 gates, d) Rabi oscillation at a Rabi frequency of 5 MHz with a \(T^{Rabi}\) of 6.6 \(\mu s\)}
\end{figure*}
In this article, we demonstrate a high-fidelity single electron spin qubit in a natural Si-MOS quantum dot device, fabricated in a 300 mm silicon on insulator (SOI) wafer. \cite{9371956, Elsayed2024-io} We perform single qubit control by sending an RF control pulse to a nearby antenna. Despite a short $T^{Ramsey} \approx 500~ \mathrm{ns}$, we achieve a Q-factor of 55 for a Rabi oscillation at a Rabi frequency of 5 MHz and a \(T^{Rabi}\) of 11 \(\mu s\) through real-time feedback which allows us to match the microwave drive and qubit frequencies. After an optimized pulse calibration protocol, a control fidelity of $99.5\pm0.3\%$ is measured through randomized benchmarking. To the best of our knowledge this is the highest fidelity single qubit operation for ESR control in natural silicon-MOS device and for n-type natural silicon-MOS structures in general. \cite{Stano2021-ch} Lastly, we study the low frequency fluctuations in the qubit frequency. 

\section{Device and experimental set-up}

Devices are fabricated on 300 mm isotopically natural SOI wafers (88 nm Si on 145 nm BOX), where the Si channel was thinned down to 10 nm by subsequent steps of dry oxidation and wet etching, followed by 50 nm epitaxial regrowth of Si. The final substrate is therefore 60 nm Si on 145 nm SiO2 (BOX). Source/drain junctions are made by implantation and subsequent anneal. The gate stack is 12 nm of dry thermal oxide and 30 nm of highly doped Poly Si. An overlapping gates architecture is used, as described in references \cite{9371956, Elsayed2024-io}. Gate layer 1 forms the confinement gate, gate layer 2 the plunger gates, and gate layer 3 the barrier gates. The barrier gate fills the gap between the plunger gates, such that the gap between them is defined by the oxide separating them. For these devices the oxide separating the gates was 5 nm of high temperature oxide. Gates are patterned using e-beam lithography and plasma etching.

Figure \ref{fig:fig1}a shows a scanning electron micrograph (SEM) of a double quantum dot device, similar to the one used in this study. The red dot indicates the single quantum dot formed under the P1 gate for the single electron qubit. The B1 gate controls the tunneling barrier from the quantum dot to the extended electron reservoir under the P2, B2, and R gates. The DC current through a single electron transistor, formed by the RB, LB, and ST gates, is used to measure charge occupancy of the quantum dot. The single electron transistor (SET) current is amplified using a Basel Precision Instruments I to V converter (SP983c) at 24 kHz bandwidth and  sampled by the Quantum Machines OPX+. An on-chip aluminium antenna similar to design (d) of reference \cite{Dehollain2013-sz} (not visible in the SEM image) is fabricated near the device to perform ESR on the electron spin qubit. An external magnetic field of \(B_{ext} = 520\) mT is applied in the z-direction. The spin state of the qubit is read out using energy-selective tunneling into the reservoir, as proposed by Elzerman et al. \cite{Elzerman2004-pc, Keith2019-zw, Mills2022-yi}

\begin{figure*}[ht]
\includegraphics[width=0.97\linewidth]{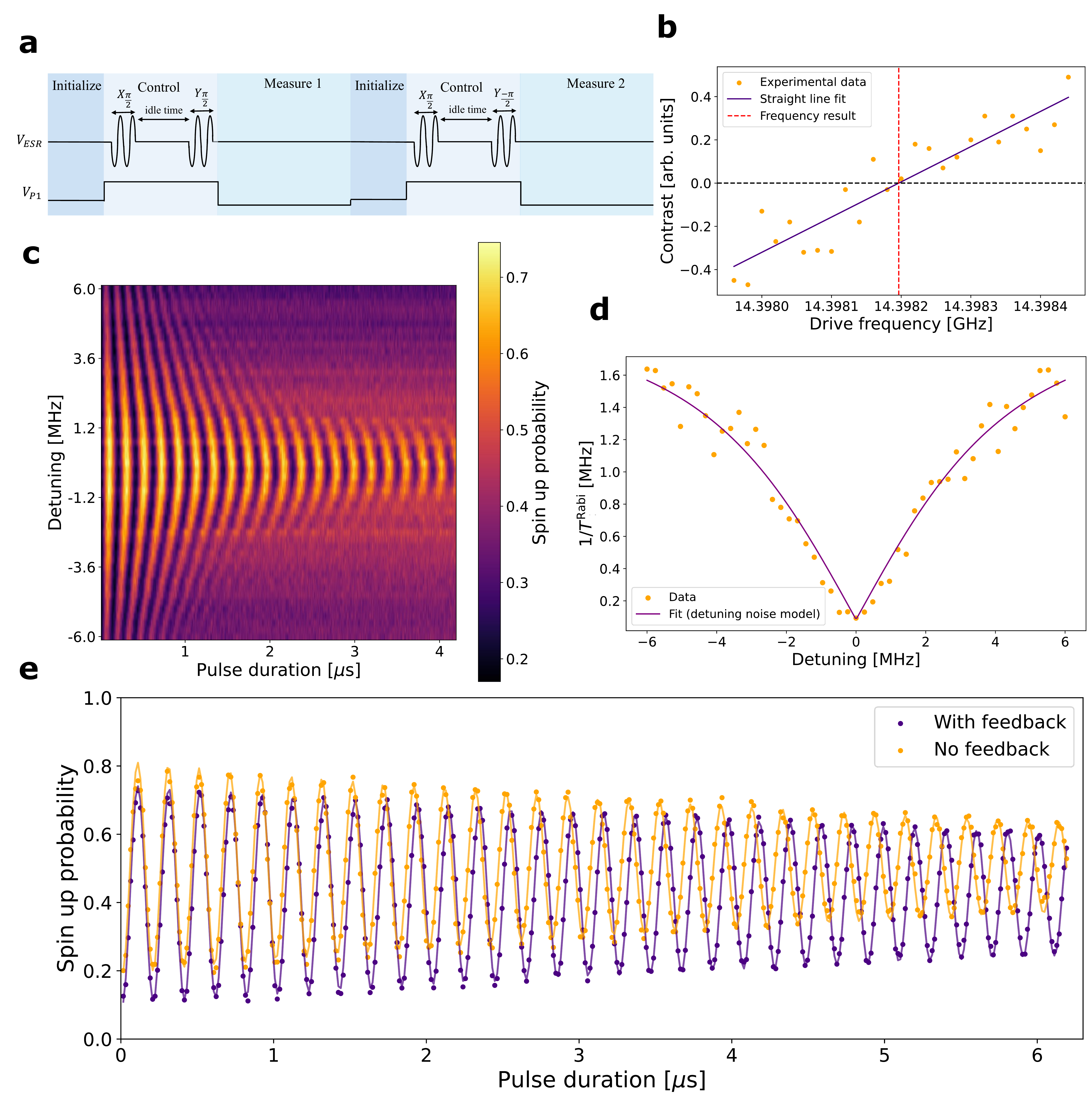}
\caption{\label{fig:fig2} \textbf{Single qubit control using qubit frequency feedback} (a) The pulse sequence for the calibration of the qubit frequency. (b) Example of the contrast signal (\(Measure_2- Measure_1\)) around resonance, with a straight line fit to locate zero detuning. (c) Rabi chevron at a Rabi frequency of 5 MHz.  (d) \(1/{T^{Rabi}}\) as a function of detuning from the qubit frequency. The line is a fit to Eq. \ref{eq:T2rabi} with a detuning uncertainty of 0.4 MHz. Hence $1/T^{Rabi}$ can be described by the uncertainty in the effective Rabi frequency. At zero-detuning, the effective Rabi frequency is a minimum and the quality factor of the Rabi oscillation is limited by the accuracy of setting zero detuning. (e) Rabi oscillations at 5 MHz on resonance, with and without feedback during the measurement. Note that the feedback reduces the Rabi frequency, and improves $T^{Rabi}$ from 7 \(\mu s\)  to 11 \(\mu s\). }
\end{figure*}

The device is held at base temperature of 10 mK in an Oxford Instruments Proteox MX dilution fridge. The electron temperature is approximately 100 mK. The RF control pulse is applied by Quantum Machines Octave. A CuBe UT085 coaxial cable with 9 dB of attenuation delivers the microwaves to a Quantum Machines QDevil 1 motherboard, with custom daughterboard. The device is bonded with aluminium wires of 17.5 $\mu m$ diameter. The Elzerman readout places a lower limit on the qubit frequency, but we found better quality control at carrier frequency of 14 GHz, than at 17.5 GHz.

To control and measure the qubit, we use the pulse sequence shown in Figure \ref{fig:fig1}b. During the control stage, the single electron spin states are pushed into the single electron regime, well below the Fermi level of the reservoir, as indicated in Figure \ref{fig:fig1}c. The initialize, control, and measure steps last 100 \(\mu s\), 300 \(\mu s\), and 600 \(\mu s\), respectively. By sending microwave pulses to the ESR antenna, an oscillating magnetic field is applied perpendicular to the externally applied magnetic field to drive Rabi oscillations between the spin-up and spin-down states. An example of a Rabi oscillation measured without feedback on the qubit frequency (ESR frequency) is shown in Figure \ref{fig:fig1}d. A fit to a single exponential envelope of a decaying oscillation yields a Rabi frequency (\(f_{Rabi}\)) of 5 MHz, a \(T^{Rabi}\) of 7 \(\mu s\), and $Q_{Rabi}=f_{Rabi}T^{Rabi}=35$.

\section{Feedback}

\begin{figure*}[!ht]
\includegraphics[width=0.95\linewidth]{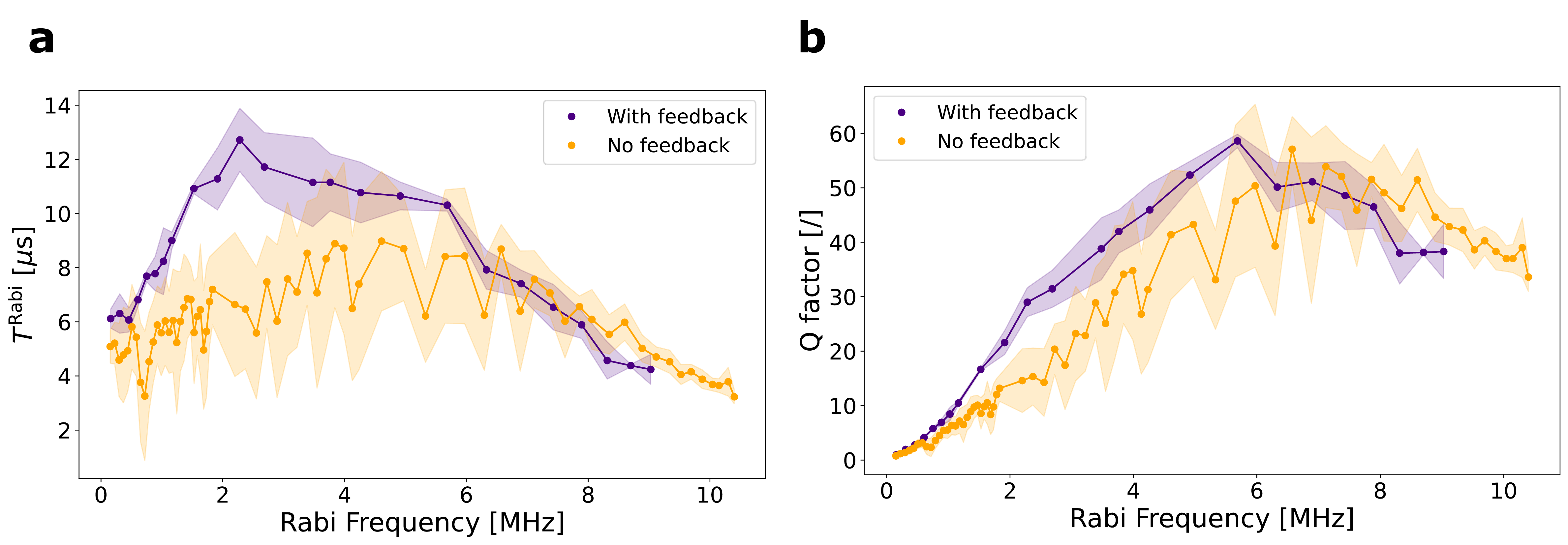}
\caption{\label{fig:fig3}\textbf{Power dependence of Rabi damping} a) \(T^{Rabi}\) as a function of Rabi frequency to study the microwave power dependence. b) Q-factor (\(Q_{Rabi}=f_{Rabi}T^{Rabi}\)) as a function of Rabi frequency. The shaded regions show one standard deviation around the average of 10 measurements of a Rabi oscillation of 600 shots. The results with and without feedback are compared and show that feedback improves the $T^{Rabi}$ for Rabi frequencies below 6 MHz, and gives more consistent measurements.}
\end{figure*}

For spin qubits, the main source of error is caused by fluctuations in the qubit frequency due to the nuclear spin, and charge environment. For low frequency noise, slow compared to the repetition rate of the measurement, the noise can be considered as a quasi-static error in the detuning that can be corrected using fast measurements of the qubit frequency \cite{Nakajima2020-cr}. Our feedback scheme is inspired by the feedback protocols described by Dumoulin Stuyck et al. \cite{Dumoulin_Stuyck2024-ql} The loop starts with a coarse measurement of the qubit frequency. To do this, a Rabi chevron  is recorded, and the peak of the Fourier transform is used to locate the zero detuning condition. Next, the measurement loop uses a Ramsey sequence to make a fast and more accurate measurement of the zero detuning condition. The drive frequency is then updated to be on resonance, followed by M Rabi oscillation measurements. This loop is repeated L times to achieve MxL averages. 

The qubit frequency measurement uses two interleaved Ramsey pulse sequences as shown in Figure \ref{fig:fig2}a. The interleaved sequence cancels low frequency variations in state preparation and measurement (SPAM) errors. In the first sequence, a \(X_{\pi/2}\) pulse is followed by an idle duration of 20 ns and a \(Y_{\pi/2}\) pulse. The second sequence ends with a \(Y_{-\pi/2}\) pulse instead. The contrast signal is calculated as the difference between the two measurements, and is proportional to the error in detuning. 
An example of the contrast signal versus drive frequency is shown in Figure \ref{fig:fig2}b. The x-intercept of a line fit locates the zero-detuning condition. The frequency range of the quasi-static noise that can be compensated is limited by the speed of the feedback. The protocol is explained in more detail in Appendix B. 

Figure \ref{fig:fig2}c shows a Rabi chevron measured using a feedback step, every 10 seconds of measurement. The Rabi decoherence rate (\(1/{T^{Rabi}}\)) as a function of detuning is extracted from the Rabi chevron shown in Figure \ref{fig:fig2}d. The data is fitted to a model expressed by:
\begin{equation}
\frac{\sqrt{2}}{T^{Rabi}_{\Delta}} \approx \frac{|\Delta|}{\sqrt{\Delta^2 + \Omega^2}} \cdot \sigma_\Delta 
+ \frac{1}{2} \cdot (\frac{\Omega^2}{(\Delta^2 + \Omega^2)^{3/2}}) \cdot \sigma_\Delta^2
\label{eq:T2rabi}
\end{equation}

Here, \(\Delta\) is the detuning, \(\Omega\) is the Rabi frequency, and \(\sigma_\Delta\) is the standard deviation of the detuning, and is the only fitting parameter. The model assumes $1/T^{Rabi}$ is proportional to the uncertainty in the effective Rabi frequency due to uncertainty in the detuning.
The full derivation is explained in Appendix B. The steep drop in the Rabi decoherence rate at zero detuning illustrates the importance of maintaining zero detuning. The fit returns an estimated detuning uncertainty of $\sigma_{\Delta} = 0.4~\mathrm{MHz}$, which is compatible with the measured uncertainty in the qubit frequency from Figure \ref{fig:fig5}. The improvement in driven coherence is illustrated in Figure \ref{fig:fig2}e, comparing the Rabi oscillations with and without feedback at a Rabi frequency of 5 MHz. We note that the Rabi frequency with feedback is slightly slower, indicating that the mean detuning is closer to zero.

\begin{figure*}
\includegraphics[width=\linewidth]{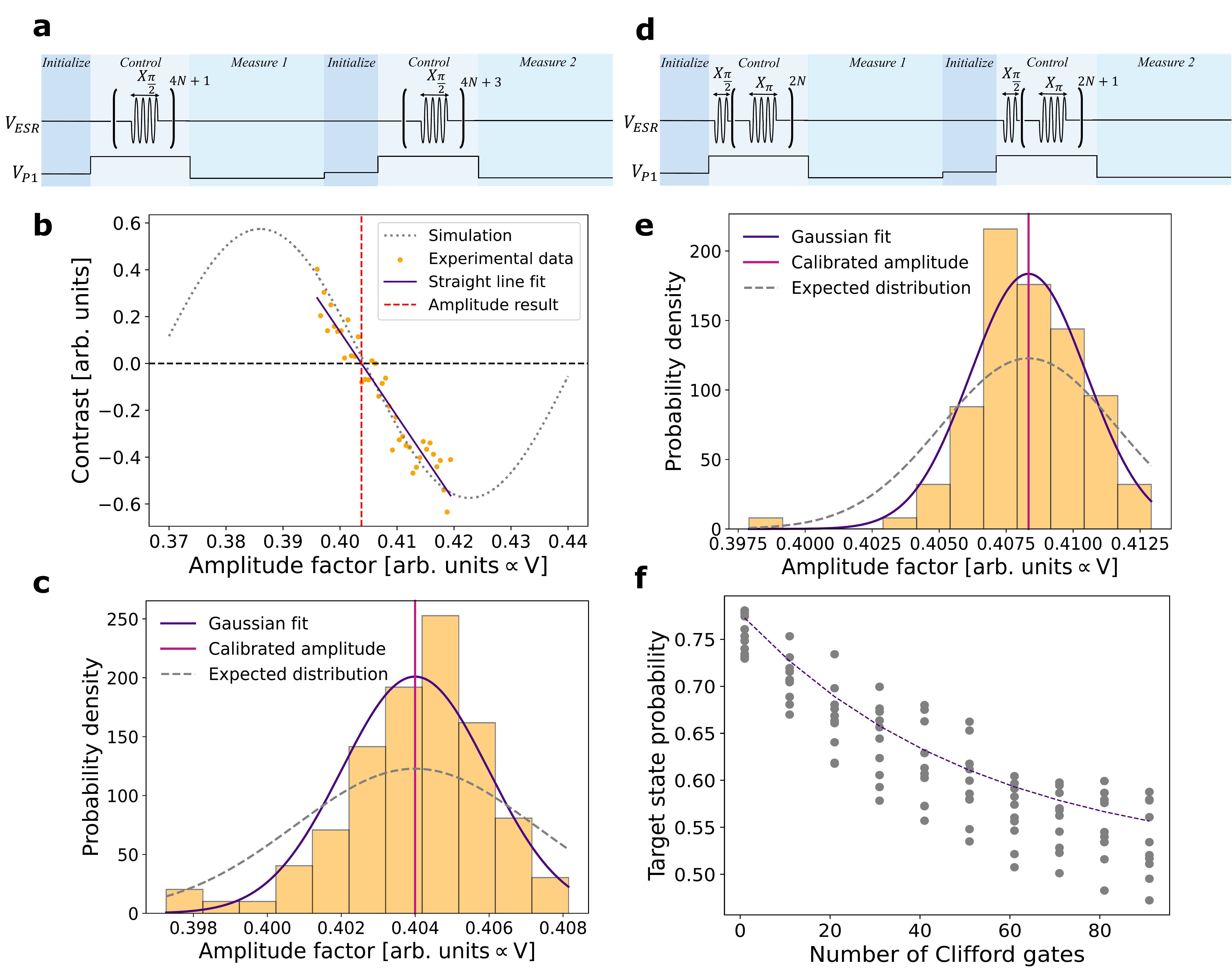}
\caption{\label{fig:fig4}\textbf{Pulse calibration and randomized benchmarking} (a) Pulse sequence for \(\pi/2\) pulse calibration. (b) An example of the contrast signal (\(Measure_2- Measure_1\)) from the \(\pi/2\) pulse calibration experiment around the matching pulse amplitude with a straight line fit to locate the optimal pulse amplitude. (c) Distribution of the optimized pulse amplitudes for a \(\pi/2\) with a gaussian fit in purple. In (c,e) The mean is shown as the violet line, and the expected pulse amplitude distribution is shown as a grey dotted gaussian based on the standard deviation of the measured qubit frequency from data of Figure \ref{fig:fig5}. (d) Pulse sequence for \(\pi\) pulse calibration. (e)  Distribution of the optimized pulse amplitudes for a \(\pi\) with a gaussian fit in purple. (f) Randomized benchmarking results, using the optimized pulse amplitudes and qubit frequency feedback, with an average single gate fidelity fitted to $99.5\pm0.3\%$ fidelity. }
\end{figure*}

Figure \ref{fig:fig3}a compares the \(T^{Rabi}\) as a function of the Rabi frequency which at low power is proportional to the amplitude of the microwave pulse, see Appendix C, without and with feedback.  Each point shows the mean and the shaded region shows one standard deviation away from the average. The feedback protocol enables higher driven coherence times in the low-power regime below 6 MHz. Additionally, the feedback protocol enhances the repeatability of the measurement in the low-power regime, as the standard deviation is reduced. The rate of Rabi damping is expected to be second-order sensitive to strong fluctuations in the detuning as $\frac{\sqrt{2}}{T^{Rabi}_{\Delta}}\approx \frac{\sigma^2_{\Delta}}{2\Omega}$, with $\sigma_{\Delta}=0.4~\mathrm{MHz}$, and first order sensitive to fluctuations in the Rabi drive, as $\frac{\sqrt{2}}{T^{Rabi}_{\Omega}}\approx \sigma_{\Omega}\approx a\Omega$, with $a\approx 0.02$. This results in an optimum Rabi frequency for the Q-factor. Furthermore, since the feedback protocol corrects detuning fluctuations only, there is no improvement at Rabi frequencies above 6 MHz. At $f_{Rabi}=6 ~\mathrm{MHz}$, the gain of the IQ-mixer is starting to saturate, see Figure \ref{fig:appC}, and this results in a higher LO-leakage. Additionally, high-frequency detuning noise could limit the driven decoherence for Rabi frequencies above 6 MHz. We note that the optimum Rabi frequency of about 5 MHz is relatively high compared to other works using ESR control \cite{Koppens2006-gv}. This is attributed to the antenna design.

\section{Pulse calibration and randomized benchmarking}

\begin{figure*}
\includegraphics[width=0.95\linewidth]{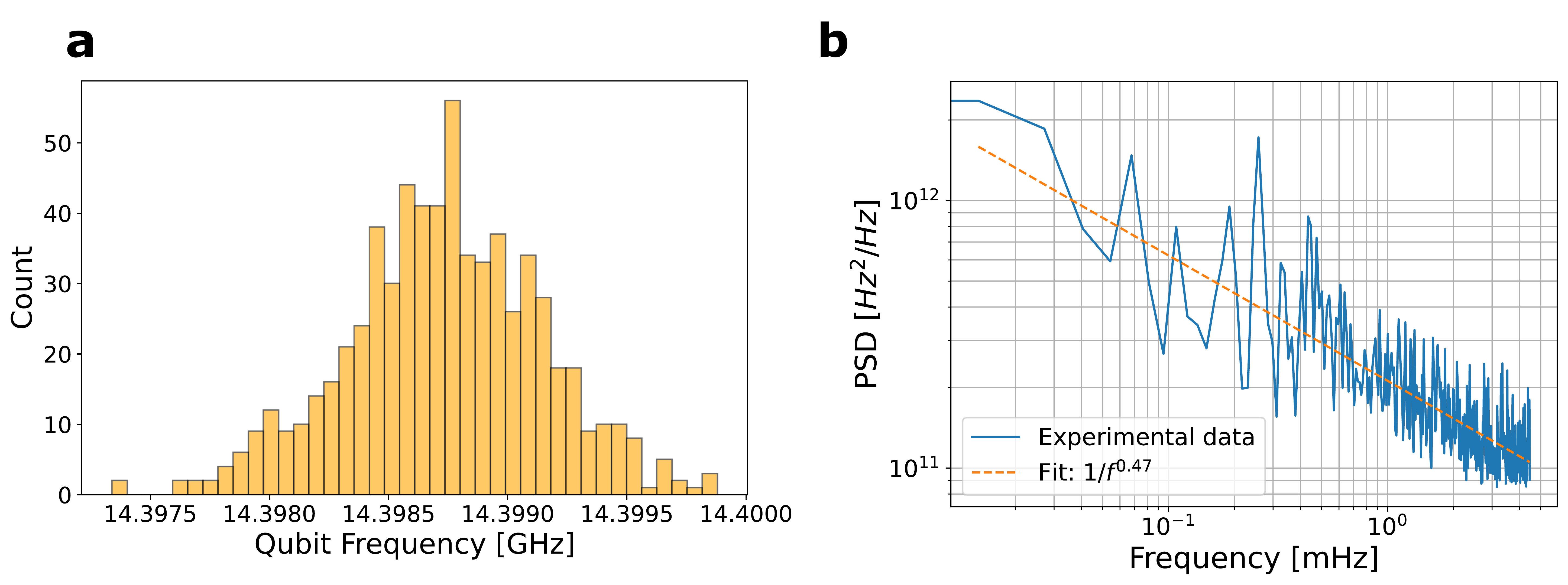}
\caption{\label{fig:fig5}\textbf{Qubit low frequency noise analysis} (a) The qubit frequency distribution measured from the qubit frequency feedback protocol during the measurement. (b) Power spectral density of the qubit frequency showing a \(1/f^{0.47}\) dependence.}
\end{figure*}

To reduce systematic pulse-area errors, we calibrate the pulse-area of the microwave pulses before doing the randomized benchmarking experiment. \cite{Dumoulin_Stuyck2024-ql} Because of the fixed 4 ns cycles of the pulse generator hardware, the pulse length is fixed and the amplitude is optimized. The \(\pi/2\) pulse calibration sequence uses two interleaved sequences of 4N + 1 and 4N + 3 \(\pi/2\) pulses, as shown in Figure \ref{fig:fig4}a. The contrast signal is the difference between the results of Measure 2 and Measure 1. This is performed for a range of pulse amplitudes. Since the contrast is proportional to the error in the pulse-area, the zero intercept of a straight line fit yields the optimum pulse amplitude. 

Figure \ref{fig:fig4}b shows an example of the contrast data and the resulting calibrated amplitude. The grey dotted line shows a simulation of the signal which oscillates with the amplitude error $\delta V$ as $\sim \sin{((4N+2)\frac{\delta V}{V})}$. The number of pulses, N, magnifies the sensitivity of the contrast to an error in the amplitude. However, increasing N, also decreases the period of the contrast signal. Therefore, N is chosen to match one half period of this oscillation to the range of possible values for the optimum amplitude. The optimization of the pulse-area calibration is detailed in appendix D. The distribution of calibrated amplitudes for the \(\pi/2\) pulse is fitted to a Gaussian distribution, as shown in Figure \ref{fig:fig4}c. This distribution is compared to the expected pulse amplitude distribution, based on the hypothesis that the uncertainty in the calibrated amplitude is limited by the noise in the qubit frequency, see appendix E. 
The mean of the distribution is chosen as the optimized amplitude. The distribution expected from variations in the detuning gives a slight overestimate of the actual distribution, which is explained in appendix E. As the measured, and expected spread in the optimum pulse amplitude are similar, the uncertainty is limited by the detuning noise during the measurement. Therefore, the feedback during the measurement is only performed on the qubit frequency, and no additional continuous feedback is performed on the pulse amplitude. 

The \(\pi\)-pulse calibration sequence contains two interleaved pulse sequences of a single \(\pi/2\)-pulse followed by 2N  and 2N+1 \(\pi\)-pulses. The mean optimum amplitude for the \(\pi \)-pulse is about \( 1\%\) higher than for  $\pi/2$-pulse, see Figure \ref{fig:fig4}e.

To measure the single gate control fidelity, a randomized benchmarking experiment \cite{PhysRevA.77.012307} is performed, see Figure \ref{fig:fig4}f. The randomized benchmarking experiment utilizes the feedback protocol every 10 seconds of measurement. The random, gapless sequences consist of pulses, drawn from the 24 Clifford gates. Each of the 24 Clifford gates is constructed by a sequence of physical pulses from the set \{\(\mathbb{I}, X_\pi, Y_\pi, X_{\pm\pi/2}, Y_{\pm\pi/2}\)\}. The average number of physical gates per Clifford gate is 1.875. \cite{Muhonen2015-rf} The identity gate is realized by leaving the qubit idle for a duration equal to that of a single \(X_\pi\) or \(Y_\pi\) pulse. For each sequence length, the measurement is repeated for $K=10$ sequences. All results are plotted, and used in the fit. The sequence fidelity decays as \(V p^m + B\) with a visibility of V=0.27, B=0.49, and a depolarizing parameter of \(1-p = 0.019 \pm 0.01\). \cite{Mills2022-yi, Gilbert2023-hh} This corresponds to a single Clifford gate infidelity of \(r_C = 0.0094 \pm 0.006\) and an average single gate infidelity of \(r_G = 0.0051 \pm 0.003\).\cite{Yoneda2018-tc} The standard deviation, skew and excess kurtosis of the fidelity are 0.3\%, -0.16 and -0.22, respectively. The spread arises from variations between different sequences due to a distribution of gate errors. This fidelity ($99.5\pm0.3\%$) exceeds the fault-tolerant threshold, despite the nuclear spin noise in the isotopically natural silicon and the Si-MOS structure.

\section{Noise analysis}

Lastly, the low-frequency noise on the qubit frequency is further analyzed. The distribution of the calibrated qubit frequencies is displayed in a histogram in Figure \ref{fig:fig5}a by sampling the qubit frequency every 2 minutes for 19 hours. The qubit frequency has a standard deviation of (\(\sigma_f = 0.41\pm0.01~MHz\)), which corresponds to a \( T_2^* = \frac{\sqrt{2}}{2\pi\sigma_f}= 560 ~\mathrm{ns}\), which compares to \( T_2^{Ramsey} \approx 500~\mathrm{ns}\). The distribution in qubit frequency is consistent with the Rabi damping data of Figure \ref{fig:fig2}d and the pulse-area calibration data of Figure \ref{fig:fig4}c and e, through the models explained in appendices B and E, respectively. If we assume the distribution in qubit frequency is due to fluctuations in the nuclear magnetic field generated by \(^{29}\mathrm{Si}\) isotopes, then by scaling the measured $T_2^*\propto r$ to calculations for Si-MOS quantum dot presented in reference \cite{Cvitkovich2024-nq}, we estimate the radius of the electron wavefunction to be  $r\approx 19~\mathrm{nm}$. The gap in the confinement gate is $w_c\approx 26~\mathrm{nm}$, and the width of the P1 gate, less the overlap with the confinement gate is $w_{P1}\approx50~\mathrm{nm}$. The $\sqrt{w_{P1}w_c} \approx 36~\mathrm{nm}$, which is close to the electron wavefunction diameter estimated from $T_2^*$.  
The time-correlated behavior of the qubit frequency is also clear in the power spectral density (PSD) of the calibrated qubit frequency, which shows a \(\frac{1}{f^{0.47}}\) dependence, see Figure \ref{fig:fig5}b. 
Due to this time-correlated behavior, the single-gate fidelity could be further enhanced by speeding up the feedback and measurement protocol. The $\frac{1}{f^{0.47}}$ behavior could be compatible with the low-frequency regime of a nuclear diffusion model or a Lorentzian indicative of a small number of two-level fluctuators. \cite{Rojas-Arias2024-zt} 

\section{Discussion}
We report single qubit gate fidelities of $99.5\pm0.3\%$. This is remarkably high for a single electron spin in a natural Si-MOS device using ESR. \cite{Stano2021-ch, Koppens2006-gv} We attribute the high gate fidelity to the high quality of the devices, resulting in low charge noise. 
In addition, we observe an optimum Rabi frequency of about 5 MHz, which is unusually high for ESR experiments. \cite{Koppens2006-gv, Huang2019-ro, Veldhorst2014-la} By accessing a regime where the Rabi frequency exceeds the spread in qubit frequencies due to the fluctuating \(^{29}\mathrm{Si}\) nuclei, the electron spin is dynamically decoupled from its environment. The high Rabi frequency is attributed to the antenna design combined with delivery of high rf-power with relatively low noise. We estimate the conversion factor of the Rabi frequency to the power applied to the PCB to be $C =0.97\pm 0.04~ \mathrm{GHz/\sqrt{W}}$. 
To further improve the Quality factor of the Rabi oscillations from 40 to 55, a feedback scheme is used to minimize the detuning between the microwave drive and the qubit frequency.  Currently, the improvement in Q-factor is limited by a slow feedback loop where the majority of the time is spent on calibration. This could be improved by streamlining the calibration, and using faster readout method such as RF-reflectometry \cite{Vigneau2023-fv}. Furthermore, a method to calibrate the pulse-area prior to randomized benchmarking is used to achieve a gate fidelity of $99.5\pm0.3\%$ that is close to the expected limit based on Rabi Q-factor: $F_{limit}\approx 1- \frac{1}{4Q_{Rabi}}=99.54\% \approx F_{measured}$. \cite{Stano2021-ch}


\section{Appendices}
\appendix
\section{Feedback protocol optimization}

The calibration protocol for the qubit frequency feedback is optimized for fast, robust, and accurate calibration. The protocol is divided into two scans: a coarse scan measuring a Rabi chevron with wide frequency range, and a fast, fine modified Ramsey sequence to better locate the qubit frequency. Both scans are optimized for accuracy and speed. The total feedback protocol takes 50 seconds to complete, consisting of the coarse scan of 40 seconds and the fine scan of 10 seconds.

For robustness, a coarse Rabi chevron is measured. A fast Fourier transform is performed on the data to select the drive frequency with the highest intensity in frequency space. The frequency range is defined by the typical distribution of the qubit frequency over time and covers a range of 2 MHz. The ESR microwave pulses are applied at low power to obtain a sharp Rabi chevron in the frequency direction. In this case, the visible width in the frequency direction is limited by the noise of the qubit frequency rather than by the Rabi frequency. The pulse duration scan is optimized to provide a fast but accurate estimate of the qubit frequency. The step size of the pulse duration is defined by the maximum step size that satisfies the Nyquist-Shannon sampling theorem for the Rabi frequency at the selected drive power, $2f_{Rabi}t_{step}=1$. The range of the pulse duration scan is defined by the decoherence time at the selected drive power to maximally discriminate between drive frequencies. To optimize the speed, the number of shots for the measurement is minimized until the accuracy degrades.

\begin{figure}
\includegraphics[width=0.95\linewidth]{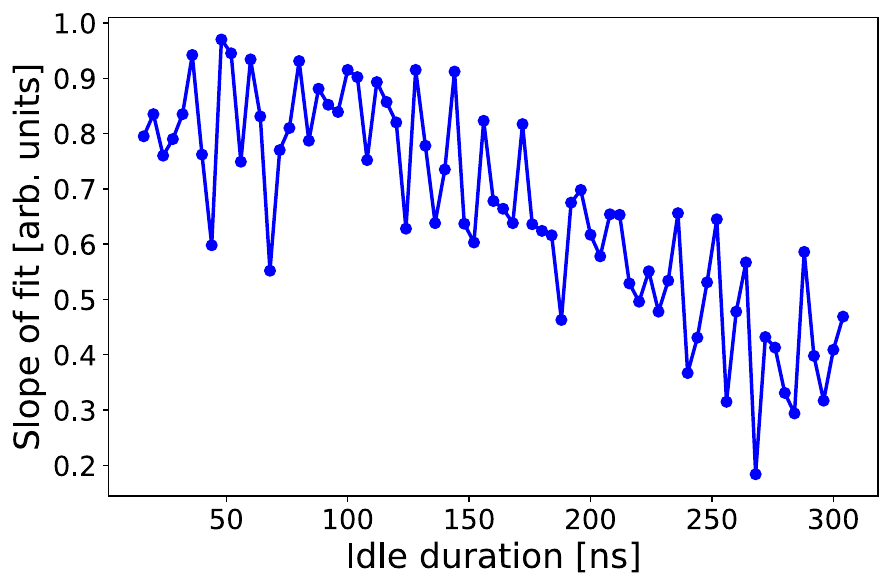}
\caption{\label{fig:appA}\textbf{Optimization of the calibration measurement} The slope of the straight line fit in Figure \ref{fig:fig2}b as a function of the idle time defined in Figure \ref{fig:fig2}a, indicating optimal sensitivity of the contrast signal to a detuning error at idle durations below 100 ns}
\end{figure}

The modified Ramsey sequence as explained in \ref{fig:fig2}a and b is used as a fine scan to provide an accurate and fast measurement of the qubit frequency. The interleaved \(X_{\pi/2}\) and \(\pm Y_{\pi/2}\) pulses are chosen such that the projection onto the z-axis is most sensitive to a detuning error near resonance. The difference or contrast in the signals  is used as the error signal as the SPAM errors are canceled. The range of the frequency scan is chosen, based on the typical noise in the qubit frequency during the modified Ramsey scan. To increase the sensitivity of  the contrast signal to a detuning error, the slope of the contrast signal, shown in Figure \ref{fig:fig2}b, is optimized using the idle time. The results are shown in Figure \ref{fig:appA} and indicate an optimal idle duration in the range of 16 to 40 ns, which is due to the short \(T_2^*\) decoherence time in this device. The drive power is reduced until the minima and maxima are visible in the defined frequency range of the scan to obtain a high slope for a robust measurement. The number of shots per frequency (30) and the frequency step size are optimized to find a balance between the speed and the accuracy of the calibration.

\section{ \(T^{Rabi}\) vs. detuning}
To model the \(T^{Rabi}\), derived from the fit to Eq. \ref{eq:T2rabi}, as a function of the drive frequency detuning (\(\Delta\)), a quasi-static decoherence model is used. This treats the detuning as static over the timescale of a measurement, but in a randomized state at the start of each measurement. The calculation of the spin-up population averages a detuned Rabi oscillation over a spread in the detuning.
\begin{equation}
P_1(t) = \int_{-\infty}^{\infty}
\frac{\Omega^2}{\Omega_{eff}^2}
\sin^2\left( \frac{t}{2} \Omega_{eff} \right)
\frac{1}{\sqrt{2\pi} \sigma} \, e^{-\frac{\Delta^2}{2\sigma^2}} \, d\Delta
\label{eq:integral}
\end{equation}
Where \( \sigma_\Delta \) is the standard deviation of the detuning distribution, the detuning mean is approximated as 0, and \(t\) is the pulse duration.

In this model, the decoherence arises from a standard deviation in the effective Rabi frequency, \(\Omega_{eff} = \sqrt{\Omega^2+\Delta^2}\). To convert the spread in detuning into a spread in effective Rabi frequency, a second-order approximation of the spread of the effective Rabi frequency \(\sigma_{\Omega_{\text{eff}}}\) is used.

\begin{equation}
\label{eq:rabi_eff}
\sigma_{\Omega_{\text{eff}}} \approx 
\frac{d\Omega_{\text{eff}}}{d\Delta} \cdot \sigma_\Delta 
+ \frac{1}{2} \cdot \frac{d^2\Omega_{\text{eff}}}{d\Delta^2} \cdot \sigma_\Delta^2
\end{equation}

With:
\begin{equation}
\frac{d\Omega_{\text{eff}}}{d\Delta} = \frac{|\Delta|}{\sqrt{\Delta^2 + \Omega^2}}
\end{equation}
\begin{equation}
\frac{d^2\Omega_{\text{eff}}}{d\Delta^2} = \frac{\Omega^2}{\left( \Delta^2 + \Omega^2 \right)^{3/2}}
\end{equation}

To convert this \(\sigma_{\Omega_{\text{eff}}}\) into the experimental \(T^{Rabi}\) that is obtained from a fit to an exponentially decaying sine wave, the integral from Eq. \ref{eq:integral} is calculated using the constant-phase approximation:

\begin{equation}
    P(t) \propto cos(\Omega_{eff} * t)e^{\frac{-1}{2}\sigma_{\Omega_{\text{eff}}}^2t^2}
\end{equation}

The exponential decay from the experimental fit, used to obtain \(T^{Rabi}\) throughout the paper, is then approximated by equating the gaussian decay from equation 4 to an exponential decay at a characteristic time \(t=T^{Rabi}\), which results in the following.

\begin{equation}
    e^{-t/T^{Rabi}}=e^{\frac{-1}{2}\sigma_{\Omega_{\text{eff}}}^2t^2}
\end{equation}
\begin{equation}
    T^{Rabi} \approx \frac{\sqrt{2}}{\sigma_{\Omega_{\text{eff}}}}
\end{equation}

\section{ Rabi frequency vs pulse amplitude}

\begin{figure}
    \centering
    \includegraphics[width=0.95\linewidth]{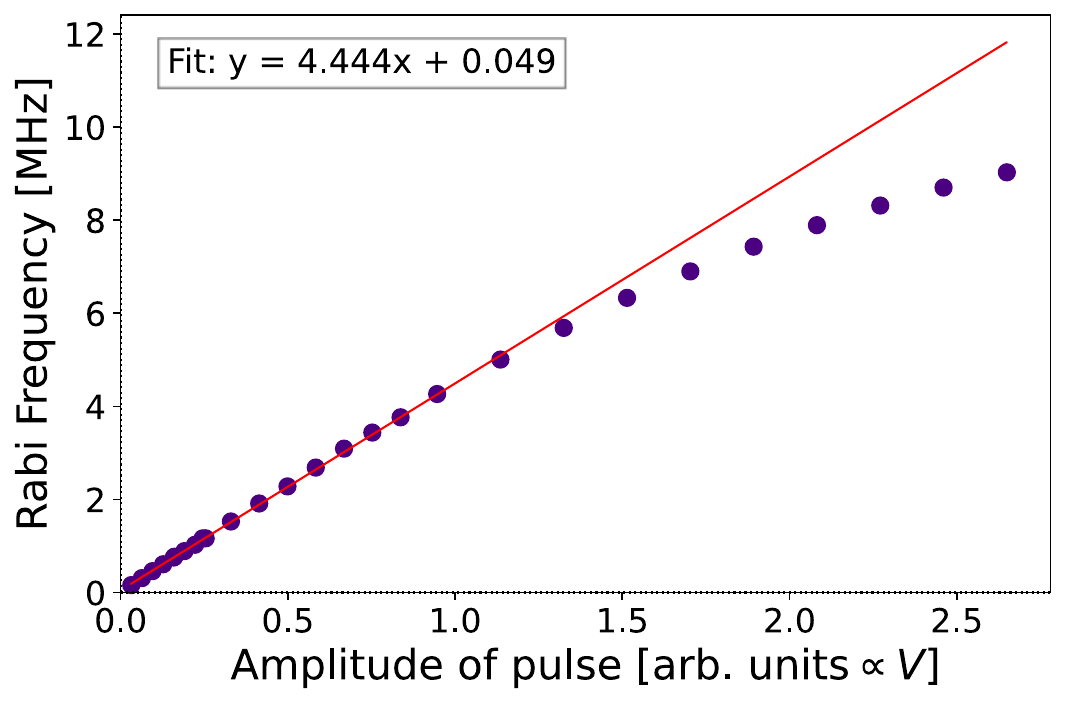}
    \caption{\textbf{Rabi frequency power dependence} The linear fit is performed on the data points with a Rabi frequency below 5 MHz}
    \label{fig:rabi_f}
\end{figure}

Figure \ref{fig:rabi_f} shows the power dependence of the Rabi frequency. The x-axis is plotted in arbitrary units proportional to the pulse amplitude voltage. A linear fit is obtained for the data points with a Rabi frequency below 5 MHz.

\section{ Pulse-area calibration protocol optimization}

Before making the randomized benchmarking measurement, the pulse-amplitude is optimized to minimize systematic pulse-area errors. Since we hypothesize that the noise in calibrated amplitudes is a result of the noise in the qubit frequency, the protocol is not integrated into the measurement as continuous feedback. Therefore, the protocol is not optimized for speed.

Since the 4-ns time-step of the AWG is coarse, the time-duration of a $\pi/2$ pulse is fixed at 52 ns. The randomized benchmarking uses a gapless sequence constructed from a set of 24 gates chosen with equal probability. This includes an identity operator, where the drive is switched off for 104 ns. Note that this idle time is about 20\% of  $T_2^{Ramsey}\approx500~\mathrm{ns}$. The pulse-area is calibrated using the amplitudes which can be defined with 16-bit resolution.

As shown in Figure \ref{fig:fig4}, an interleaved sequence of 4N+1 and 4N+3 $\pi/2$-pulses is applied. As before, the contrast signal of the interleaved pulses cancels low frequency SPAM errors and is maximally sensitive to the pulse-area. The sensitivity to pulse-area is magnified by using N pulses.

\begin{figure}
\includegraphics[width=0.95\linewidth]{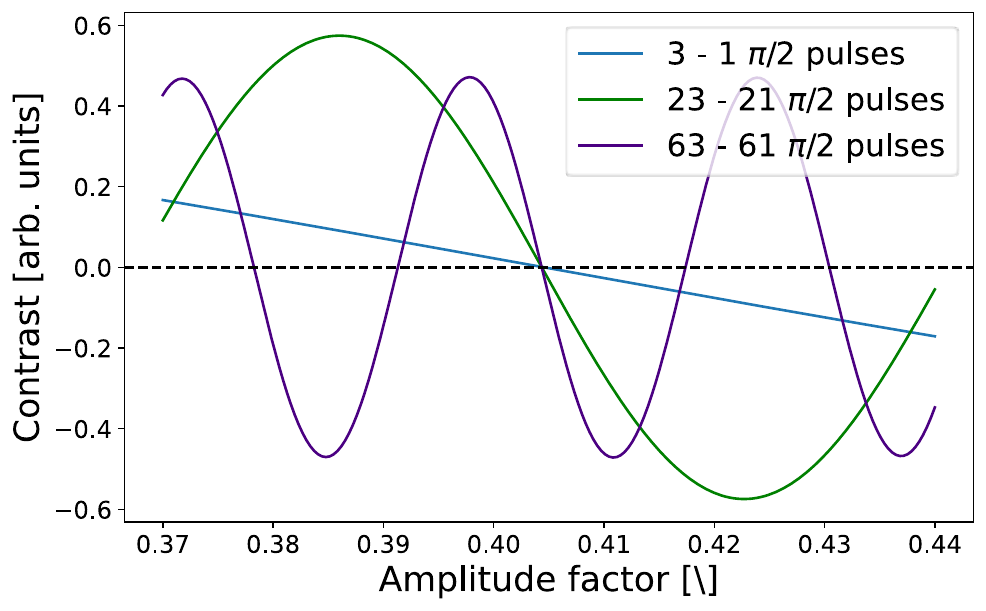}
\caption{\label{fig:appC}\textbf{Optimization of the pulse-area calibration} Simulation of the pulse sequence in Figure \ref{fig:fig4}a for different \(N\) values, as defined in Figure \ref{fig:fig4}a. The figure indicates the dependence of the \(N\) parameter on the sensitivity of the contrast signal to an amplitude error around the matching condition}
\end{figure}

The amplitude range of the scan is first defined by the typical noise in the calibrated amplitudes. The number of pulses (as a parameter N) is then increased to obtain the highest sensitivity (slope), around the matching condition between the pulse length and amplitude, until the amplitude range contains the minima and maxima of the contrast signal. This can be seen in Figure \ref{fig:fig4}b where the range for the frequency scan of the experiment is chosen to be between the minima and maxima of the simulated contrast signal. Since the period of the oscillations in amplitude are defined by N, as shown in Figure \ref{fig:appC}, the slope is maximized by increasing the number of pulses, N. In the case of a short driven decoherence time \(T^{Rabi}\) or large N, the decoherence may impact the optimal N, as the contrast signal is reduced when many pulses are applied over a timescale comparable to the decoherence time.


\begin{figure*}
\includegraphics[width=0.9\linewidth]{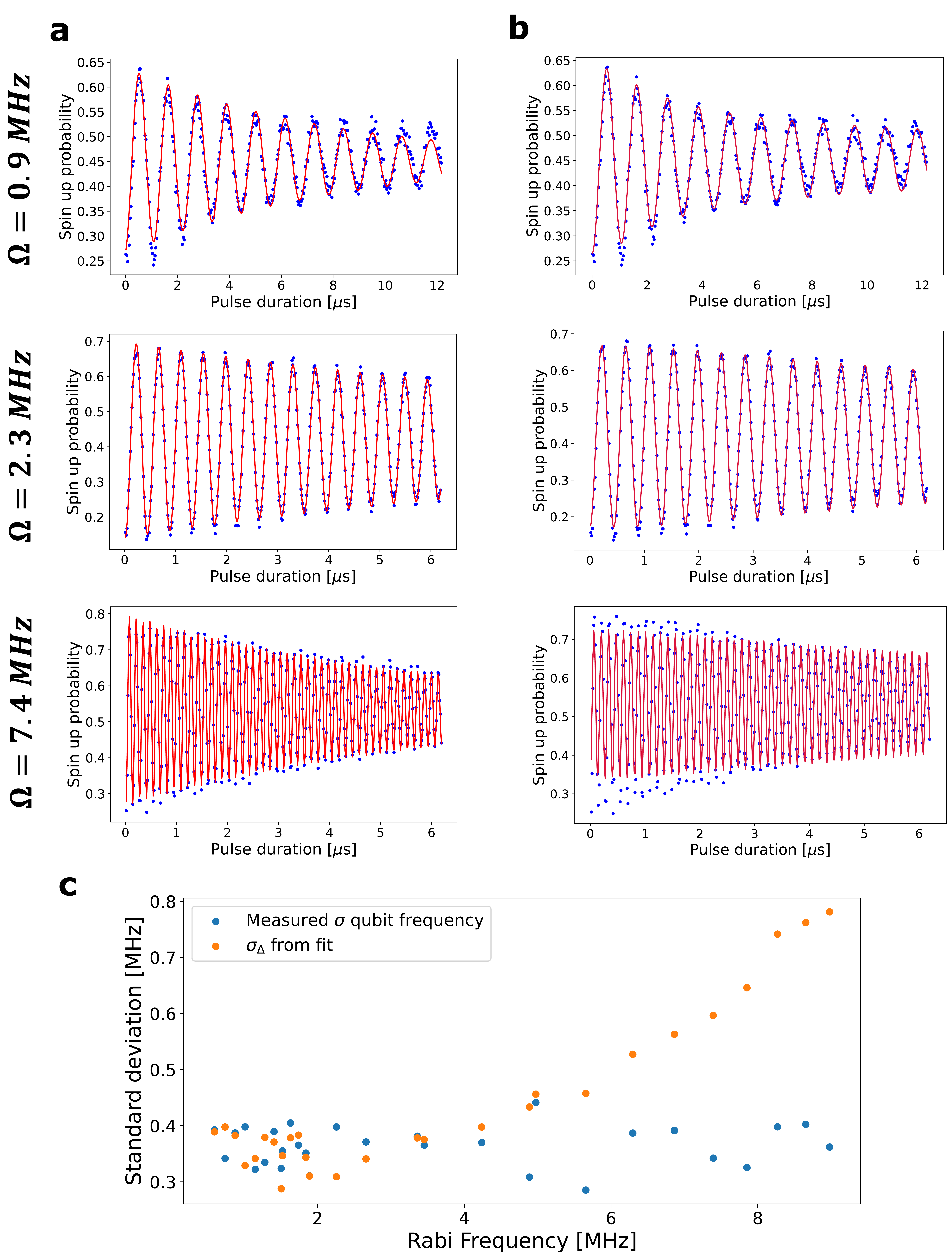}
\caption{\label{fig:oscillationfit}\textbf{Analysis of Rabi oscillations decay} The Rabi oscillations are fit to different models to study the limiting decoherence factors, showing the experimental data in blue and the fitted function in red. (a) The Rabi oscillations with an exponentially decaying cosine oscillation fit for different RF pulse powers. (b) The Rabi oscillations fitted to the model described by \(P_{model, 1}(t)\), taking into account a gaussian distribution in the detuning and fitting for the standard deviation of this detuning distribution. (c) Comparison of uncertainty in qubit frequency vs Rabi frequency measured using: (blue) spread in qubit frequencies, (orange) fit to Rabi oscillation data. }
\end{figure*}

\section{Expected pulse amplitude distribution derivation}

The expected standard deviation of the amplitude factor is derived based on the standard deviation of the effective Rabi frequency due to detuning noise. A second-order approximation of the effective Rabi frequency is made as a function of a change in detuning, as derived in \ref{eq:rabi_eff}. This function is evaluated for \(\Delta=0\), which is an approximation of the operating point of this measurement.

\begin{equation}
    \sigma_{\Omega_{\text{eff}}} \approx 
\frac{1}{2\cdot \Omega} \cdot \sigma_\Delta^2
\end{equation}

The standard deviation in the effective Rabi frequency is then converted into the standard deviation of the pulse amplitude factor by using the slope (\(b\)) of the amplitude dependence of the Rabi frequency as fitted in Figure \ref{fig:rabi_f} to obtain the following result.

\begin{equation}
    \sigma_{amplitude} \approx 
\frac{1}{2\cdot \Omega \cdot b} \cdot \sigma_\Delta^2
\end{equation}

The \(\sigma_\Delta\) is the standard deviation of the detuning during the measurement. This is approximated as \(0.38 \pm 0.01~MHz\) which is derived from the distribution of the consecutive differences between the calibrated qubit frequencies. However, since these calibrations are performed at a slower rate than the rate at which the measurement is performed after the calibration, the effective detuning distribution during the measurement is expected to have a smaller standard deviation due to the time-correlated noise in the qubit frequency. This approximation is thus expected to be an overestimate.

\section{Analysis of Rabi oscillations decay}
Additional analysis is performed on the Rabi oscillation decay fit to gain insight into the mechanism behind the decoherence. To obtain a measure of the driven decoherence time \(T^{Rabi}\), an exponentially damped sine wave is fitted to the data. This fit is compared with the fit of the model described in Appendix B, a decoherence model that only considers a spread in detuning (\(\Delta\) with standard deviation \(\sigma\) and mean 0 Hz) over the duration of the measurement for a constant Rabi frequency (\(\Omega\)).

Figure \ref{fig:oscillationfit}b shows the results from fitting the Rabi oscillations to the following model, where \(t\) is defined as the pulse duration. 


\begin{eqnarray}
P_1(t) = \int_{-\infty}^{\infty}
\frac{\Omega^2}{\Omega_{eff}^2} \cdot
\sin^2\left( \frac{\Omega_{eff}t}{2}  \right) \cdot g(\Delta,\sigma) d\Delta \nonumber \\
g(\Delta,\sigma) = \frac{1}{\sqrt{2\pi} \sigma} \, e^{-\frac{\Delta^2}{2\sigma^2}} 
\end{eqnarray}

The full model uses the fitting parameters A and C, to accommodate for the SPAM errors in the experimental data.

\begin{equation}
P_{\text{model, 1}}(t) = A \left( P_1(t) - \frac{1}{2} \right) + C
\end{equation}






Qualitatively, the decoherence model fits the Rabi oscillations at low Rabi frequencies (low drive power). At low Rabi frequencies, the model follows the Rabi oscillation at longer pulse durations more accurately than the conventional exponential decay. Additionally, the detuning standard deviation can be derived by fitting the decoherence model to the oscillation. This obtained standard deviation from the Rabi oscillations is compared with the measured qubit frequency standard deviation at different drive powers in Figure \ref{fig:oscillationfit}c. The model returns a good estimate of the detuning standard deviation at low Rabi frequencies below 6 MHz, but fails to obtain a good fit of the Rabi oscillation at higher drive powers.

We hypothesize that the decoherence model provides an accurate description of the  decoherence for low drive powers with a Rabi frequency below 6 MHz. At higher drive powers,  the driven decoherence rate is limited by high frequency noise, and a more exponential-like decay.

\section*{Acknowledgements}
This work was financially supported by JST Moonshot R\&D under Grant Number JPMJMS2065, by the Engineering and Physical Sciences Research Council (EPSRC)
under Grant Number EP/S022953/1. X.P. acknowledges the University of Cambridge Harding Distinguished Postgraduate 
Scholars Programme. This work was also supported by the Royal Society through a University Research Fellowship held by H.S.K.

\section*{REFERENCES}

\bibliography{apssamp}


\end{document}